\documentclass[fleqn,usenatbib]{mnras}
\usepackage{newtxtext,newtxmath}
\usepackage[T1]{fontenc}
\DeclareRobustCommand{\VAN}[3]{#2}
\let\VANthebibliography\thebibliography
\def\thebibliography{\DeclareRobustCommand{\VAN}[3]{##3}\VANthebibliography}
\usepackage{graphicx}	
\usepackage{amsmath}
\usepackage{lineno}
\usepackage{color}
\usepackage{graphicx}
\usepackage{longtable}
\usepackage{multirow}
\usepackage{url}
\usepackage{subfigure}
\usepackage{lipsum}
\usepackage{ulem}
\usepackage{bm}

\title[An 8\% Determination of $H_0$ from FRBs]{An 8\% Determination of the Hubble Constant from localized Fast Radio Bursts}

\author[Q. Wu et al.]{
Qin Wu,$^{1}$
Guo-Qiang Zhang,$^{1}$
Fa-Yin Wang$^{1,2}$ \thanks{E-mail: fayinwang@nju.edu.cn}
\\
$^{1}$School of Astronomy and Space Science, Nanjing University, Nanjing 210093, China\\
$^{2}$Key Laboratory of Modern Astronomy and Astrophysics (Nanjing University), Ministry of Education, Nanjing 210093, China\\
}

\date{Accepted XXX. Received YYY; in original form ZZZ}
\pubyear{2022}

\begin{document}
\label{firstpage}
\pagerange{\pageref{firstpage}--\pageref{lastpage}}
\maketitle

\begin{abstract}
The cosmological-constant ($\Lambda$) cold dark matter (CDM) model is challenged by the Hubble tension, a remarkable difference of Hubble constant $H_0$ between measurements from local probes and the prediction from Planck cosmic microwave background observations under $ \Lambda$CDM model.
So one urgently needs new distance indicators to test the Hubble tension. 
Fast radio bursts (FRBs) are millisecond-duration pulses occurring at cosmological distances, which are attractive cosmological probes. 
Here we report a measurement of ${H_0}  = 68.81^{+4.99}_{-4.33} {\rm \ km \ s^{-1} \ Mpc^{-1}}$ using eighteen localized FRBs, with an uncertainty of 8\% at 68.3 per cent confidence.
Using a simulation of 100 localized FRBs, we find that error of $H_0$ can be reduced to 2.6\% at $1\sigma$ uncertainty. 
Thanks to the high event rate of FRBs and localization capability of radio telescopes (i.e., Australian Square Kilometre Array Pathfinder and Very Large Array), future observations of a reasonably sized sample will provide a new way of measuring $ H_0$ with a high precision to test the Hubble tension.
\end{abstract}


\section{Introduction}

The cosmological-constant ($\Lambda$) cold dark matter (CDM) model successfully explains the majority of cosmological observations \citep{Planck20}. 
The value of Hubble constant ($H_0$), describing the expansion rate of our universe, is a basic and fascinating issue in cosmology.
The measurements of cosmic microwave background (CMB) by Planck Collaboration \citep{Planck20} are powerful probes to estimate cosmological parameters. 
The distance-redshift relation of specific stars (e.g. Cepheid variables and type Ia supernovae) can be used constrain $H_0$ directly \citep{Riess20,Riess21}. 
Other methods are also used to measure $H_0$, such as Baryon Acoustic Oscillations (BAO), gravitational lensing \citep{Wong2020} and Gravitational Waves (GWs) \citep{Abbott17}. 
Great advances in modern observational technology have improved the precision of measuring $H_0$. 
However, a significant difference at least 4$\sigma$ is reflected in the Hubble constant measured by CMB and Cepheid-calibrated type Ia supernovae (SNe Ia) respectively, known as “Hubble tension” \citep{Freedman17,DiValentino21}.  
New physics or observational bias are two mainstream arguments to alleviate this tension. 
An independent and robust method of measuring $H_0$ should be used to test this tension. 

Fast radio bursts (FRBs) are short-duration radio pulses with enormous dispersion measures (DMs) \citep{Lorimer07,Katz18,Petroff19,Cordes19,Platts19,Xiao21}.
In a short period of more than ten years, a vigorous development has appeared in the observation of FRBs.
Up to now, more than 600 FRBs have been observed, including repeating FRBs.
There are 19 FRBs with definite host galaxies and redshift measurements.
The host galaxy association and redshift measurement point out the direction for the study of the origin, radiation mechanism and cosmological application of FRBs. 

Dispersion Measure (DM) is defined as the integral of the number density of free electrons along the propagation path, which is positively proportional to cosmological distance. 
In particular, $\rm DM_{IGM}$, contributed by the intergalactic medium (IGM), has a close connection with cosmological parameters. 
Therefore, precise measurements of $\rm DM_{IGM}$ can be used as cosmological probes \citep{Xiao21,Bhandari2021}, such as "missing" baryons \citep{McQuinn14,Macquart20,Li2020}, cosmic proper distance \citep{Yu17}, dark energy \citep{Zhou14,Walters18,Zhao2020,Qiu2021}, Hubble parameter $H(z)$ \citep{Wu2020}, and the cosmic reionization history \citep{Zheng14,Zhangzj21}. 
Strongly lensed FRBs have been proposed to probe the nature of dark matter \citep{Munoz16,Wang2018}, and measure Hubble constant \citep{Li18}. 

There is a thorny problem that DMs contributed by host galaxy and the inhomogeneities of intergalactic medium cannot be exactly determined from observations \citep{Macquart20}. 
Previous works assuming fixed values for them bring uncontrolled systematic error in analysis \citep{Hagstotz21}. A reasonable approach is to handle them as probability distributions extracted from cosmological simulations \citep{Macquart20,Jaroszynski19,Zhang20,Zhangzj21}. 

In this letter, we propose to measure Hubble constant with eighteen localized FRBs through the $\rm DM_{IGM}$-$z$ relation.
Our paper is organized as the following four sections. 
In section \ref{sec2}, we present the redshift and DM value of eighteen localized FRBs used in our analysis.
In section \ref{sec3}, we give an introduction of the distributions of $\rm DM_{host}$ and $\rm DM_{IGM}$. 
In section \ref{sec4}, the Monte Carlo Markov Chain (MCMC) analysis is used to constrain the Hubble constant $H_0$.  
Discussion will be given in section \ref{sec5}.

\section{The properties of localized FRBs}\label{sec2}

\begin{table}
	\small
	\begin{center}
		\caption{ Properties of localized FRBs.}
		\begin{tabular}{ccccc}
			\hline
			Name & Redshift & DM$_{\rm obs}$   & Reference \\
			&  & $(\rm pc \ cm^{-3})$  & &  \\
			\hline
			FRB 121102 &    0.19273 &$557\pm 2$	    & \cite{Chatterjee17}\\
			FRB 180301 &    0.3304  &$522\pm 0.2$           & \cite{Bhandari21}\\
			FRB 180916 &	0.0337  &$349.349\pm 0.005$	      & \cite{Marcote20}\\
			FRB 180924 &    0.3214  &$361.42\pm 0.06$        & \cite{Bannister19}\\
			FRB 181030 &    0.0039  &$103.396\pm 0.005$      & \cite{Bhardwaj21B}\\
			FRB 181112 &    0.4755	&$ 589.27\pm 0.03$      & \cite{Prochaska19}\\
			FRB 190102 &    0.291 	&$ 363.6\pm 0.3$	       & \cite{Bhandari20}	\\
			FRB 190523 &    0.66    &$ 760.8\pm 0.6$        & \cite{Ravi19}\\
			FRB 190608 &    0.1178  &$ 338.7\pm 0.5$	       & \cite{Chittidi20}	\\
			FRB 190611 & 	0.378   &$ 321.4\pm 0.2$	     & \cite{Heintz20}  \\
			FRB 190614 &    0.6     &$ 959.2\pm 0.5$            & \cite{Law20}\\
			FRB 190711 &    0.522 	&$ 593.1\pm 0.4$     & \cite{Heintz20}\\
			FRB 190714 &    0.2365	&$ 504 \pm 2$         & \cite{Heintz20}\\
			FRB 191001 &	0.234   &$ 506.92\pm 0.04$        & \cite{Heintz20}\\
			FRB 191228 &    0.2432  &$ 297.5\pm 0.05$         & \cite{Bhandari21}\\
			FRB 200430 &    0.16	&$ 380.1\pm 0.4$      & \cite{Heintz20}\\
			FRB 200906 &    0.3688  &$ 577.8\pm 0.02$          & \cite{Bhandari21}\\
			FRB 201124 & 	0.098   &$ 413.52\pm 0.05$      & \cite{Ravi21}\\
			\hline
		\end{tabular}
		\label{data}
		\vspace{0.5cm}
	\end{center}
\end{table}

A remarkable feature of FRB is that its DM value is much larger than that contributed by the Milky way. 
The DM$_{\rm obs}$ obtained directly from observations can be divided into the following components:
\begin{equation}
\rm{DM_{obs}}(z) =  \rm{DM_{MW} + DM_{IGM}}(z) + \frac{\rm{DM_{host}}(z)}{1+z}, 
\label{DMobs}
\end{equation}
where $\rm DM_{MW}$ is contributed by the interstellar medium (ISM) and the halo of the Milky Way, $\rm DM_{IGM}$ represents contribution from the IGM, $\rm DM_{host}$ is the contribution by the host galaxy. It is necessary to consider the value of each term in equation (\ref{DMobs}) separately. 
$\rm DM_{MW}$ can be separated into the ISM-contributed $\rm DM_{MW, ISM}$ and the halo-contributed $\rm DM_{MW, halo}$. 
The NE2001 model is used to derive $\rm DM_{MW, ISM}$ \citep{Cordes02}. 
This model estimates DM contributions from the galaxy ISM with the orientation of the Galactic-coordinate grids. 
For the halo-contributed $\rm DM_{MW,halo}$, it has been estimated that the Galactic halo will contribute $ 50 \sim 80\ \rm pc \ cm^{-3}$ from the Sun to 200 kpc \citep{Prochaska2019}. 
Here we assume a Gaussian distribution with a mean value of $\rm 65\ pc \ cm^{-3}$ and a standard deviation of $\rm{15 \ pc \ cm^{-3}}$ as the probability distribution of $\rm DM_{MW,halo}$ to consider the uncertainty of $\rm DM_{MW,halo}$.

For $\rm DM_{IGM}$, the effect of IGM inhomogeneities will lead to significant sightline-to-sightline scatter around the mean $\rm DM_{IGM}$ \citep{McQuinn14}. 
The scatter of DM at $z=1$ is about 400 pc cm$^{-3}$ from theoretical analysis \citep{McQuinn14} and the state-of-the-art cosmological simulations \citep{Jaroszynski19,Zhangzj21}.
Considering a flat universe, the averaged value of $\rm DM_{IGM}$ is \citep{Deng14}
\begin{equation}\label{dmigm}
\langle {\rm{DM_{IGM}}(z)}\rangle = \frac{A \Omega_bH_0^2}{H_0}\int_{0}^{z_{\rm FRB}}\frac{f_{\rm IGM}(z)f_e(z)(1+z)}{\sqrt{\Omega_m(1+z)^3+ 1 - \Omega_m}}dz, 
\end{equation}
where $ A = \frac{3c}{ 8\pi G m_p}$ and $m_p$ is the proton mass. The electron fraction is $f_e(z) = Y_H X_{e, H}(z) + \frac{1}{2} Y_{He}X_{e, He}(z)$, with hydrogen fration
$Y_H  = 0.75$ and helium fraction $Y_{He} = 0.25$. 
Hydrogen and helium are completely ionized at $z<3$, which implies the ionization fractions of intergalactic hydrogen and helium $X_{e,H} = X_{e,He} = 1$. 
The cosmological parameters $\Omega_b$ and $\Omega_m$ are the the density of baryons and the density of matter, respectively.
At present, there is no observation that can give the evolution of the fraction of baryon in the IGM $f_{\rm IGM}$ with redshift.  \cite{Shull12} gave an estimation of $f_{\rm IGM} \approx 0.83$. 

Currently, 19 FRBs have been localized including the nearest repeating FRB 200110E \citep{Bhardwaj21}, which is located in a globular cluster in the direction of the M81 galaxy \citep{Kirsten21}. 
The distance of FRB 200110E is only 3.6 Mpc, which is the closest-known extragalactic FRB so far. 
Correspondingly, the DM value of FRB 200110E is 87.75 $\rm pc \ cm^{-3}$.
And the intergalactic medium (IGM) contributed $\rm DM_{IGM} = 1 pc \ cm^{-3}$ is estimated from the relation of the averaged $\rm DM_{IGM}$ and redshift. 
Thus the cosmological information carried by FRB 200110E is too seldom to calculate $H_0$. Additionally, the effect of peculiar velocity cannot be ignored, which makes it trick to calculate cosmological parameters. Therefore, we excluded FRB 200110E from the localized FRB sample. 
In general, a sample with a larger amount of data will be more accurate to constrain parameters by reducing the statistical error. 
Thus we choose the other eighteen localized FRBs to constrain the Hubble constant.
Table \ref{data} shows the redshifts, DM$_{\rm obs}$, $\rm DM_{MW,ISM}$, telescopes and references of eighteen localized FRBs.

The $\rm DM_{IGM}$ value of FRBs can be estimated using $\rm{DM_{IGM}} = \rm{DM_{obs}} - \rm{DM_{MW,ISM} - DM_{MW,halo} - DM_{host}}/(1+z)$. 
Here the value of $\rm DM_{MW,ISM}$ is estimated from the NE2001 model \citep{Cordes02} with the FRB coordinates.  We also test our results using the YMW16 model \citep{Yao17}, and find the effect can be neglected for different free electron distribution models.
A Gaussian distribution is used to describe the probability distribution of $\rm DM_{MW,halo}$.

\begin{figure}
	\centering
	\includegraphics[width=\linewidth]{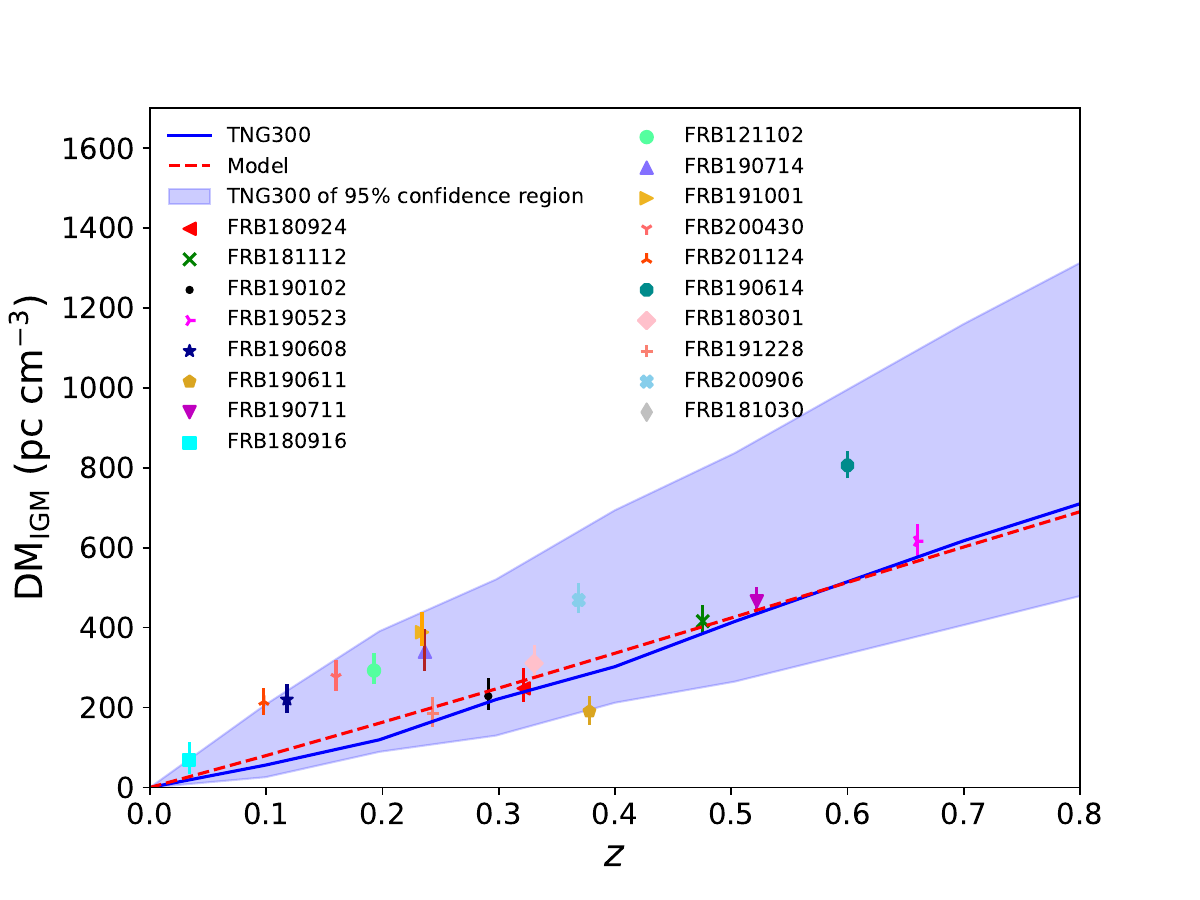}
	\caption{{\bf The $ \rm \bf DM_{IGM}$-$\bm z$ relation for eighteen localized FRBs.} 
		The scattered points are the $\rm DM_{IGM}$ values of the eighteen localized FRBs. The $\rm DM_{IGM}$ values are derived by correcting the observed dispersion measure $\rm DM_{obs}$ for the estimated contributions from our Galaxy and the host galaxy. The $\rm DM_{MW,ISM}$ is deduced from NE2001 model, and $\rm DM_{MW,halo}$ is adopted as a Gaussian distribution with a median of 65 pc cm$^{-3}$. We use the median value of $\rm DM_{host}$ at different redshifts from IllustrisTNG 300 cosmological simulation \citep{Zhang20}. The uncertainty of $\rm DM_{IGM}$ is estimated from equation (\ref{error}). The red dotted line shows model of equation (\ref{dmigm}) with $\Omega_{\rm m} = 0.315$, $\Omega_{\rm B}h^2 = 0.02235$ and $H_0 = {\rm 70\ km\ s^{-1}\ Mpc^{-1}}$. The blue line corresponds to the $\rm DM_{IGM}$ result from the IllustrisTNG 300 cosmological simulation and the purple shaded area is the 95$\%$ confidence region \citep{Zhangzj21}. Apparently, some FRBs significantly deviates from the averaged $\rm DM_{IGM}$ by considering the median value of $\rm DM_{host}$. Therefore, in order to obtain reliable cosmological constraints, the probability distributions of $\rm DM_{IGM}$ and $\rm DM_{host}$ must be considered.
	}
	\label{fig_data}
\end{figure}

According to equation (\ref{DMobs}), the uncertainty of $\rm DM_{IGM}$ can be estimated as:
\begin{equation}\label{error}
\sigma_{\rm IGM}(z) = \sqrt{\sigma_{\rm obs}(z)^2 + \sigma^2_{\rm MW} + \left( \frac{\sigma_{\rm host}(z)}{1 + z} \right)^2},
\end{equation}
where $\sigma_{\rm obs}$ is the uncertainty of $\rm DM_{obs}$. $\sigma_{\rm MW}\approx {\rm 30\ pc/cm^{-3}} $ is the sum of the uncertainty of $\rm DM_{MW,halo}$ and $\rm DM_{MW,ISM}$. And $\sigma_{\rm host}$ is the uncertainty of $\rm DM_{host}$. The evolution of the median of $\rm DM_{host}$ can be fitted by ${\rm DM_{host}}(z) = A(1+z)^{\alpha}$, where $A$ and $\alpha$ are given in \cite{Zhang20}. The uncertainty of $\rm DM_{host}$ comes from the uncertainties of $A$ and $\alpha$. { As given in \cite{Zhang20}, $A$ and $\alpha$ have upper limits and lower limits. The maximum value of $\rm DM_{host}$ is calculated from the maximum value of $A$ and $\alpha$. Similarly, the minimum value of $\rm DM_{host}$ can be calculated. And the difference between the minimum value and the center value of $\rm DM_{host}$ and the difference between the maximum value and the center value of $\rm DM_{host}$ are the uncertainties of $\rm DM_{host}$. }
The $\rm DM_{host}$ is adopted as the median value derived from the IllustrisTNG simulation \citep{Zhang20}.
Therefore, ${\rm DM_{IGM}}$ can be estimated by subtracting the above terms and the ${\rm DM_{IGM}}$-$z$ relation of these FRBs is shown in Figure \ref{fig_data}. 
The estimated $\rm DM_{IGM}$ of eighteen localized FRBs are shown as scatters. The red dotted line is the averaged value of $\rm DM_{IGM}$ from the equation (\ref{dmigm}). The error bar gives the uncertainty of $\rm DM_{IGM}$ using the equation (\ref{error}). The blue solid line is the $\rm DM_{IGM}$ derived from the IllustrisTNG simulation with 95\% confidence region (blue shaded area) \citep{Zhangzj21}. 

\section{The distributions of $\rm DM_{host}$ and $\rm DM_{IGM}$}
\label{sec3} 

The electron number density along different sightlights is not uniform while clustering and fluctuating, so it is difficult to determine the real value of $\rm DM_{IGM}$. 
A quasi-Gaussian function with a long tail was used to fit the probability distribution of $\rm DM_{IGM}$ \citep{McQuinn14}.
This model includes scatter in the electron distribution, which is mainly caused by random variation of halos along a given sightline. 
Cosmological simulations indicate that this variation is dominated by galactic feedback redistributing baryons around galactic halos. 
Strong feedback can expel baryons to larger radii from their host galaxies. 
This form of $\rm DM_{IGM}$ distribution combines the effect of large-scale structure associated with voids and the sightlines intersecting with clusters. 
This physical-motivated model of $\rm DM_{IGM}$ distribution provides a successful fit of a wide range of cosmological simulations \citep{Macquart20,Zhangzj21}. 
Using the state-of-the-art IllustrisTNG cosmological simulation \citep{Springel18}, \cite{Zhangzj21} realistically estimated the distribution of $\rm DM_{ IGM}$ at different redshifts. 
We use the best-fit values of parameters for $\rm DM_{IGM}$ distribution at different redshifts refer to their papers.  

The $\rm DM_{IGM}$ can be fitted by a Gaussian distribution \citep{McQuinn14},
\begin{equation}\label{PIGM}
P_{\rm IGM}(\Delta) = A \Delta^{-\beta} \exp  \left[-\frac{(\Delta^{-\alpha}-C_0)}{2\alpha^2\sigma^2_{\rm DM}} \right], \Delta > 0, 
\end{equation}
where $\Delta \equiv \rm DM_{IGM}/\langle DM_{IGM}\rangle$. The indices $\alpha$ and $\beta$ are related to the inner density profile of gas in halos. 
\cite{Macquart20} gave the best fit of $\alpha =3$ and $\beta = 3$. 
$\sigma_{\rm DM}$ is an effective standard deviation. $C_0$ is a free parameter, which can be fitted when the averaged $\langle \Delta \rangle = 1$. 
The fitting values $A$, $C_0$ and $\sigma_{\rm DM}$ refer to the results of the state-of-the-art IllustrisTNG simulation \citep{Zhangzj21}.

The distribution of $\rm DM_{host}$ can be well expressed with a log-normal distribution \citep{Macquart20,Zhang20}
\begin{equation}\label{Phost}
P({\rm DM_{host}}; \mu, \sigma_{\rm host}) = \frac{1}{{\rm DM_{host}} \sigma_{\rm host} \sqrt{2\pi}} {{\exp}}\left(-\frac{{\rm ln} {\rm DM_{host}}-\mu}{2\sigma_{\rm host}^2}\right),
\end{equation}
where $e^{\mu}$ and $e^{2\mu+\sigma_{\rm host}^2}(e^{\sigma_{\rm host}^2}-1)$ are the mean and variance of the distribution, respectively. 
The distribution of $\rm DM_{host}$ derived from state-of-the-art IllustrisTNG simulation with different properties of galaxies describes the $\rm DM_{host}$ well \citep{Zhang20,Jaroszynski20}. 
\cite{Zhang20} estimated the $\rm DM_{host}$ distribution of repeating FRBs like FRB 121102, repeating FRBs like FRB 180916 and non-repeating FRBs individually. 
The redshift evolution of $\rm DM_{host}$ is also considered \citep{Zhang20}. 
Here we divide the localized FRBs into three types according to the properties of host galaxy. 

\cite{Prochaska2019} give an estimation of $\rm{ DM_{MW, halo} \approx 50 - 80 \ pc \ cm^{-3}} $. Based on their estimations, we consider a Gaussian distribution to describe the distribution of $\rm{ DM_{MW, halo}}$:
\begin{equation}\label{Phalo}
    P({\rm DM_{MW,halo}}; \mu_{\rm halo}, \sigma_{\rm halo}) = \frac{1}{\rm \sigma_{\rm halo} \sqrt{2\pi}} {{\exp}}\left(-\frac{ {\rm DM_{MW, halo}}-\mu_{\rm halo}}{2\sigma_{\rm halo}^2}\right),
\end{equation}
where we assume the mean value $\mu_{\rm halo} = 65 \ {\rm pc \ cm^{-3}}$ and the standard deviation $\sigma_{\rm halo} = 15 \ {\rm pc \ cm^{-3}}$.

We estimate the likelihood function by calculating the joint likelihoods of eighteen FRBs
\begin{equation}\label{likelihood}
    {\mathcal{L} = \prod\limits_{i=1}^{N_{\rm FRB}} P_{i}\left(\mathrm{DM}_{\mathrm{FRB}, i}^{\prime} \right)}, 
\end{equation}
where ${P_{i}\left(\mathrm{DM}_{\mathrm{FRB}, i}^{\prime} \right)}$ is the probability of individual observed FRB with $\rm DM_{{FRB}}^{\prime}=DM_{obs}-DM_{MW,ISM}=\mathrm{DM_{host}}+DM_{IGM}+DM_{MW,halo}$. For a FRB at redshift $z_i$, we have

\begin{equation}\label{Ptotal}
\begin{split}
P_{i}\left(\mathrm{DM}_{\mathrm{FRB}, i }^{\prime} \right) 
&=  \int_{0}^{\mathrm{DM}^{\prime}_\mathrm{FRB} - \rm{DM_{MW,halo}}} \int_{50}^{80}  P_{\text {host}}({\rm DM_{host}}) \\
&\times P_{\text {IGM }}({ \mathrm{DM}_{\mathrm{FRB}, i }^{\prime}-{\rm DM_{host}} - {\rm DM_{MW,halo}}})  \\
&\times P_{\rm{halo}}({\rm DM_{MW,halo}}) d{\rm DM_{host}}d{\rm DM_{MW,halo}} ,
\end{split}
\end{equation}
where $P_{\text {\rm host}}({\rm DM_{host}})$ is the probability density function (PDF) for $\rm DM_{host}$ with a mean value $\mu$ and standard deviation $\sigma_{\rm host}$,  $P_{\text {IGM}}({\rm DM_{IGM}})$ is the PDF for $\rm DM_{IGM}$ and $P_{\text {halo}}({\rm DM_{MW,halo}})$ is the PDF for $\rm DM_{halo}$.
In the calculation, according to the properties of host galaxy, FRBs can be divided into repeating FRBs like FRB 121102, repeating FRBs like FRB 180916 and non-repeating FRBs \citep{Zhang20}.

\section{Results}\label{sec4}

\begin{figure}
	\centering
	\includegraphics[width=\linewidth]{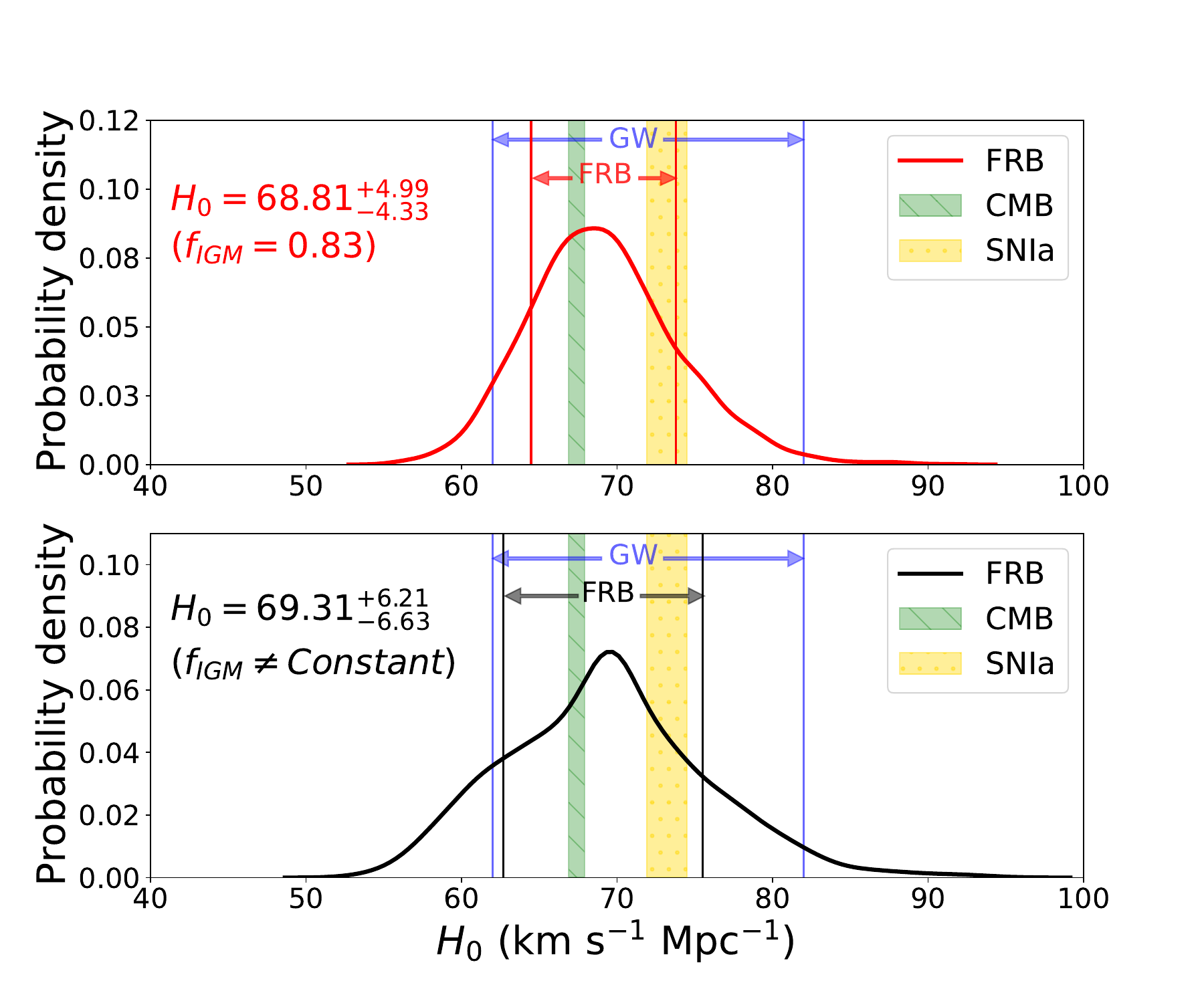}
	\caption{{ The probability density distribution of $\bm H_0$ from eighteen localized FRBs. 
	The red solid line shows the probability density distribution of $H_0$ when $f_{\rm IGM}$ is assumed as a constant 0.83 in the upper panel. The two red vertical line shows the result $H_0 = 68.81^{+4.99}_{-4.33} {\rm \ km \ s^{-1} \ Mpc^{-1}}$ with 1$\sigma$ uncertainty. 
	While the black solid line in the bottom panel shows the probability density distribution of $H_0$ when $f_{\rm IGM}$ is assumed as a variable parameter. The result is $H_0 = 69.31^{+6.21}_{-6.63} {\rm \ km \ s^{-1} \ Mpc^{-1}}$. }
	The purple vertical lines corresponds to the $H_0$ value derived by GW170817 \citep{Abbott17}. The yellow and green regions correspond to the 1$\sigma$ uncertainty range of $H_0$ reported by SH0ES \citep{Riess21} and Planck \citep{Planck20}, respectively.
	}
	\label{fig_H0}
\end{figure}

We use Monte Carlo Markov Chain (MCMC) analysis to estimate $H_0$ with eighteen localized FRBs. 
The MCMC method is based on Bayesian theory. For any prior distribution, only the properties of the required posterior distribution need to be calculated.
Current observations show that there is a general consistency for $\Omega_b h^2$ and $\Omega_m$ from different probes.
Here we consider a Gaussian distribution of $\Omega_m = 0.315\pm 0.007$ as the prior distribution of $\Omega_m$ \citep{Planck20}.  
As for $\Omega_b h^2$, it is necessary to apply an independent measurement besides CMB to break the degeneracy between $\Omega_b h^2$ and $H_0$. 
In the standard theory of Big Bang nucleosynthesis (BBN), the D/H abundance ratio has a strong relationship with the baryonic mass density. 
We use $\Omega_b h^2 = 0.02235 \pm 0.00049$ derived from the primordial deuterium abundance D/H \citep{Cooke18}, where $h = H / (100\rm{ \ km \ s^{-1} \ Mpc^{-1}})$. 
{ The fraction of baryon $f_{\rm IGM}$ in the IGM has not yet been precisely determined. 
There is also a degeneracy between $f_{\rm IGM}$ and $H_0$. The best way to break the degeneracy is to measure the value of $f_{\rm IGM}$ in future observations.
To discuss the impact of $f_{\rm IGM}$ on $H_0$, we consider two cases. 
For simplicity but without loss of generality, we assume it as a constant $f_{\rm IGM} = 0.83$ \citep{Shull12}. Meanwhile, we also consider the situation that $f_{\rm IGM}$ is not a constant, which satisfies a uniform distribution  $[0.747, 0.913]$ and an initial value $f_{\rm IGM} = 0.83$. }
The tension of $H_0$ measurements between estimation of CMB and direct model-independent measurements of supernovae in the local universe is obvious. 
We conservatively assume that the prior distribution of $H_0$ satisfies a uniform distribution in [0 - 100] km/s/Mpc.  
The distributions $\rm DM_{host}$ and $\rm DM_{IGM}$ obtained from the IllustrisTNG simulation are used \citep{Zhang20, Zhangzj21}.

\begin{figure}
	\includegraphics[width=\linewidth]{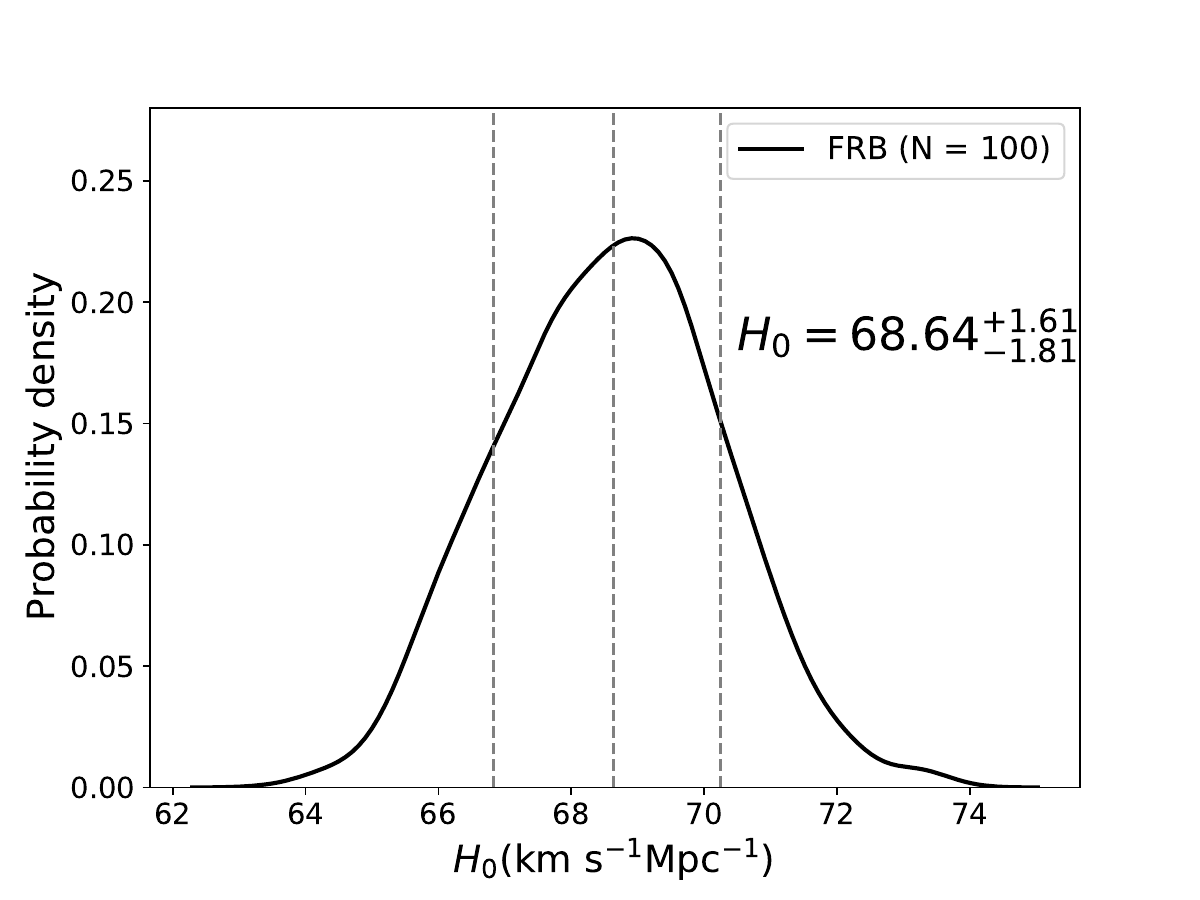}
	\caption{{\bf The probability density distribution of $\bm H_0$ from 100 simulated FRBs.}
		The black solid line corresponding to the probability density distribution of $H_0$ from 100 simulated FRBs. The dotted line in the middle corresponds to the median value of the distribution, and the dotted lines on both sides represent the 1$\sigma$ confidence interval.  
		The result is ${ H_0 =  68.64^{+1.61}_{-1.81} {\rm  \ km \ s^{-1} \ Mpc^{-1}}}$ with 1$\sigma$ error. The statistical error is about 2.6$\%$.
	} 
	\label{sim_H0}
\end{figure}

The steps of MCMC analysis are as follows: 
\begin{enumerate}
    \item The first step is to get the ($z, {\rm DM'_{FRB}}$) parameters of each localized FRBs, where $\rm DM'_{FRB}$ is estimated by subtracting the $\rm DM_{MW,ISM}$ value obtained by the NE2001 model.

    \item Equations ({\ref{PIGM}}), ({\ref{Phost}}) and ({\ref{Phalo}}) are used to model the distributions of $\rm DM_{host}$, $\rm DM_{IGM}$ and $\rm DM_{MW,halo}$ with the ($z, {\rm DM'_{FRB}}$) parameters for each localized FRB. 

    \item From the second step, $\rm{DM_{{FRB}}^{\prime}}$ is simulated by calculating the convolution of the probability density of $\rm DM_{host}$, $\rm DM_{IGM}$ and $\rm{ DM_{MW, halo}}$ to derive the probability density of $\rm DM_{FRB}^{\prime}$. As shown in equation (\ref{likelihood}), the product of the probability densities of all FRBs is a joint likelihood function. { Then MCMC method can be used to fit $H_0$, $\Omega_m$, $\Omega_b h^2$ and $f_{\rm IGM}$. Here the purpose of modelling the distributions of $\Omega_m$, $\Omega_b h^2$ and $f_{\rm IGM}$ is to consider the effect of observational errors or hypothetical errors caused by these three parameters on the result of $H_0$. }

    \item After calculating the total likelihood function, the prior distributions and the initial value of four parameters ($H_0$, $\Omega_m$, $\Omega_b h^2$ and $f_{\rm IGM}$) need to be determined. We assume a uniform prior distribution of $H_0$ in the interval $[0, 100] \rm \ km \ s^{-1} \ Mpc^{-1}$, which is a broad scope to show the properties of the posterior distribution. We suppose a initial value of $H_0 = 70 \ \rm \ km \ s^{-1} \ Mpc^{-1} $. For $\Omega_m$, we consider 1$\sigma$ error range $[0.296, 0.32]$ given by CMB as a uniform prior distribution. The initial value of $\Omega_m$ is consistent with the optimum value of measurement of CMB. Finally, we assume a uniform prior for $\Omega_b h^2$ in the interval $[0.02186, 0.02284]$, which is consistent with the 1$\sigma$ range calculated by BBN \citep{Cooke18}. The initial value of $\Omega_b h^2$ is consistent with the optimum value of BBN measurement. { And for $f_{\rm IGM}$, a fixed value 0.83 is assumed. For comparison, we also consider the case that $f_{\rm IGM}$ satisfies a uniform prior distribution $[0.747, 0.913]$. }

    \item Lastly, we run 1,000 steps of MCMC using the \texttt{emcee} package of Python \citep{Foreman-Mackey13} with the likelihood function and the priors. The plot of the final result $H_0 = 68.81^{+4.99}_{-4.33} {\rm \ km \ s^{-1} \ Mpc^{-1}}$ with 1$\sigma$ uncertainty is shown in Figure \ref{fig_H0} as red solid line for eighteen localized FRBs. This value is consistent with that derived from observed Hubble parameters $H(z)$ through the Gaussian Process method  \citep{Yu18}. The $H_0$ results with 1-$\sigma$ confidence region measured by Planck CMB data and Cepheid-based distance ladder measurement are shown as oral and yellow bands in Figure \ref{fig_H0}. They are within the 1$\sigma$ range of $H_0$ derived from FRBs. { The result $H_0 = 69.31^{+6.21}_{-6.63} {\rm \ km \ s^{-1} \ Mpc^{-1}}$ is derived when $f_{\rm IGM}$ is assumed as a variable value, which is also shown in Figure \ref{fig_H0} with black solid line. The 1$\sigma$ uncertainty with a variable $f_{\rm IGM}$ is 9.6\%, which is larger than the 1$\sigma$ uncertainty derived from the fixed $f_{\rm IGM}$. }
\end{enumerate}
It is optimistic to measure $H_0$ using a large sample of FRBs. 
Considering that a large sample of FRBs has been detected by CHIME \citep{CHIME21}, together with precise localization capability of ASKAP, VLA and Deep Synoptic Array, a sample containing 100 localized FRBs will be available in near future.
In order to predict the future measurement of $H_0$, we simulate 100 FRBs with redshifts and dispersion measures. 
Firstly, we suppose that FRBs and long gamma-ray bursts have similar redshift distribution \citep{Yu17}, which is estimated as $f(z)\propto ze^{-z}$ in the redshift $0<z<3$. 
A FRB sample with 100 mocked redshifts can be generated from the redshift distribution through Monte Carlo simulations. 
The $\rm DM'_{FRB}$ corresponding to each mocked redshifts can be obtained according to its probability distribution function. 
Then we repeat the MCMC analysis described in the previous section with simulated data.  
The simulated 100 FRBs give a result of $H_0 = 68.19^{+1.72}_{-1.02} {\rm \ km \ s^{-1} \ Mpc^{-1}}$ with an uncertainty of 2.6\% at 1$\sigma$ confidence region as shown in Figure \ref{sim_H0}. 
This precision is comparable to that $H_0$ measurement with a sample of 75 Milky Way Cepheids \citep{Riess21}. 
The result means that 100 localized FRBs can give a high-precision measurement of $H_0$, which is obviously exciting and may be realized in the near future. 

\section{Discussion}\label{sec5}

{ The statistical and systematic errors must be are discussed. 
Statistical error mainly comes from the small number of the localized FRBs. 
In order to mitigate the influence of the statistical error, we consider a sample including all the localized FRBs except FRB 200110E.
Once there are enough data, statistical error will not dominate. 
By that time, it is necessary to select data samples and classify them. 
According to the four criteria proposed by \cite{Macquart20}, we selected eight FRBs from eighteen localized FRBs. 
The statistical error is 13.4$\%$ applying the same method to constrain $H_0$. 
The main reason for the difference is the relative small sample after selection.
To explore the influence of different electron density models of the Milky Way on the final results. 
We estimate $\rm DM_{MW,ISM}$ based on the YMW16 model \citep{Yao17}, and find that the result of $H_0$ is similar as that of NE2001 model. }

{The systematic uncertainty caused by the choices of priors of $\Omega_b h^2$ and $\Omega_m$ also needs to be considered. 
Since the uncertainty of $\Omega_b h^2$ is about 2\% from BBN, we consider it is 2\%. Similarly, the uncertainty of $\Omega_m$ is about 4\%, we suppose the value of the systematic uncertainty as 4\%. 
The calculation of systematic error satisfies the error transfer formula.
We estimate that the systematic error of $H_0$ is about 4.7\%, which is smaller than the statistical error of 8\%. However, the influence of systematic error does not have a rule and can not be eliminated. 
More precise measurements of $\Omega_b h^2$ and $\Omega_m$ are needed. }

In summary, measuring Hubble constant with localized FRBs is optimistic, which supports that FRB can be treated as a reliable cosmological probe. 
FRBs can provide a new direction for solving the "Hubble tension" with more localized FRBs and precise measurements of $\Omega_b h^2$ and $\Omega_m$.

\section*{Acknowledgements}

We thank the anonymous referee for helpful comments. This work was supported by the National Natural Science Foundation of China (grant No. U1831207), and the China Manned Spaced Project (CMS-CSST-2021-A12).
We thank Z. Q. Hua for discussion. 

\section*{Data Availability}
The data that support the plots within this paper and other findings of this study are available from the corresponding author upon reasonable request.

\bibliographystyle{mnras}
\bibliography{ms} 

\begin{thebibliography}{}
\makeatletter
\relax
\def\mn@urlcharsother{\let\do\@makeother \do\$\do\&\do\#\do\^\do\_\do\%\do\~}
\def\mn@doi{\begingroup\mn@urlcharsother \@ifnextchar [ {\mn@doi@}
  {\mn@doi@[]}}
\def\mn@doi@[#1]#2{\def\@tempa{#1}\ifx\@tempa\@empty \href
  {http://dx.doi.org/#2} {doi:#2}\else \href {http://dx.doi.org/#2} {#1}\fi
  \endgroup}
\def\mn@eprint#1#2{\mn@eprint@#1:#2::\@nil}
\def\mn@eprint@arXiv#1{\href {http://arxiv.org/abs/#1} {{\tt arXiv:#1}}}
\def\mn@eprint@dblp#1{\href {http://dblp.uni-trier.de/rec/bibtex/#1.xml}
  {dblp:#1}}
\def\mn@eprint@#1:#2:#3:#4\@nil{\def\@tempa {#1}\def\@tempb {#2}\def\@tempc
  {#3}\ifx \@tempc \@empty \let \@tempc \@tempb \let \@tempb \@tempa \fi \ifx
  \@tempb \@empty \def\@tempb {arXiv}\fi \@ifundefined
  {mn@eprint@\@tempb}{\@tempb:\@tempc}{\expandafter \expandafter \csname
  mn@eprint@\@tempb\endcsname \expandafter{\@tempc}}}

\bibitem[\protect\citeauthoryear{{Abbott} et~al.,}{{Abbott}
  et~al.}{2017}]{Abbott17}
{Abbott} B.~P.,  et~al., 2017, \mn@doi [\nat] {10.1038/nature24471}, \href
  {https://ui.adsabs.harvard.edu/abs/2017Natur.551...85A} {551, 85}

\bibitem[\protect\citeauthoryear{{Amiri} et~al.,}{{Amiri}
  et~al.}{2021}]{CHIME21}
{Amiri} M.,  et~al., 2021, \mn@doi [\apjs] {10.3847/1538-4365/ac33ab}, \href
  {https://ui.adsabs.harvard.edu/abs/2021ApJS..257...59A} {257, 59}

\bibitem[\protect\citeauthoryear{{Bannister} et~al.,}{{Bannister}
  et~al.}{2019}]{Bannister19}
{Bannister} K.~W.,  et~al., 2019, \mn@doi [Science] {10.1126/science.aaw5903},
  \href {https://ui.adsabs.harvard.edu/abs/2019Sci...365..565B} {365, 565}

\bibitem[\protect\citeauthoryear{{Bhandari} \& {Flynn}}{{Bhandari} \&
  {Flynn}}{2021}]{Bhandari2021}
{Bhandari} S.,  {Flynn} C.,  2021, \mn@doi [Universe]
  {10.3390/universe7040085}, \href
  {https://ui.adsabs.harvard.edu/abs/2021Univ....7...85B} {7, 85}

\bibitem[\protect\citeauthoryear{{Bhandari} et~al.,}{{Bhandari}
  et~al.}{2020}]{Bhandari20}
{Bhandari} S.,  et~al., 2020, \mn@doi [\apjl] {10.3847/2041-8213/ab672e}, \href
  {https://ui.adsabs.harvard.edu/abs/2020ApJ...895L..37B} {895, L37}

\bibitem[\protect\citeauthoryear{{Bhandari} et~al.,}{{Bhandari}
  et~al.}{2022}]{Bhandari21}
{Bhandari} S.,  et~al., 2022, \mn@doi [\aj] {10.3847/1538-3881/ac3aec}, \href
  {https://ui.adsabs.harvard.edu/abs/2022AJ....163...69B} {163, 69}

\bibitem[\protect\citeauthoryear{{Bhardwaj} et~al.,}{{Bhardwaj}
  et~al.}{2021a}]{Bhardwaj21}
{Bhardwaj} M.,  et~al., 2021a, \mn@doi [\apjl] {10.3847/2041-8213/abeaa6},
  \href {https://ui.adsabs.harvard.edu/abs/2021ApJ...910L..18B} {910, L18}

\bibitem[\protect\citeauthoryear{{Bhardwaj} et~al.,}{{Bhardwaj}
  et~al.}{2021b}]{Bhardwaj21B}
{Bhardwaj} M.,  et~al., 2021b, \mn@doi [\apjl] {10.3847/2041-8213/ac223b},
  \href {https://ui.adsabs.harvard.edu/abs/2021ApJ...919L..24B} {919, L24}

\bibitem[\protect\citeauthoryear{{Chatterjee} et~al.,}{{Chatterjee}
  et~al.}{2017}]{Chatterjee17}
{Chatterjee} S.,  et~al., 2017, \mn@doi [\nat] {10.1038/nature20797}, \href
  {https://ui.adsabs.harvard.edu/abs/2017Natur.541...58C} {541, 58}

\bibitem[\protect\citeauthoryear{{Chittidi} et~al.,}{{Chittidi}
  et~al.}{2021}]{Chittidi20}
{Chittidi} J.~S.,  et~al., 2021, \mn@doi [\apj] {10.3847/1538-4357/ac2818},
  \href {https://ui.adsabs.harvard.edu/abs/2021ApJ...922..173C} {922, 173}

\bibitem[\protect\citeauthoryear{{Cooke}, {Pettini}  \& {Steidel}}{{Cooke}
  et~al.}{2018}]{Cooke18}
{Cooke} R.~J.,  {Pettini} M.,   {Steidel} C.~C.,  2018, \mn@doi [\apj]
  {10.3847/1538-4357/aaab53}, \href
  {https://ui.adsabs.harvard.edu/abs/2018ApJ...855..102C} {855, 102}

\bibitem[\protect\citeauthoryear{{Cordes} \& {Chatterjee}}{{Cordes} \&
  {Chatterjee}}{2019}]{Cordes19}
{Cordes} J.~M.,  {Chatterjee} S.,  2019, \mn@doi [\araa]
  {10.1146/annurev-astro-091918-104501}, \href
  {https://ui.adsabs.harvard.edu/abs/2019ARA&A..57..417C} {57, 417}

\bibitem[\protect\citeauthoryear{{Cordes} \& {Lazio}}{{Cordes} \&
  {Lazio}}{2002}]{Cordes02}
{Cordes} J.~M.,  {Lazio} T.~J.~W.,  2002, arXiv e-prints, \href
  {https://ui.adsabs.harvard.edu/abs/2002astro.ph..7156C} {pp
  astro--ph/0207156}

\bibitem[\protect\citeauthoryear{{Deng} \& {Zhang}}{{Deng} \&
  {Zhang}}{2014}]{Deng14}
{Deng} W.,  {Zhang} B.,  2014, \mn@doi [\apjl] {10.1088/2041-8205/783/2/L35},
  \href {https://ui.adsabs.harvard.edu/abs/2014ApJ...783L..35D} {783, L35}

\bibitem[\protect\citeauthoryear{{Di Valentino} et~al.,}{{Di Valentino}
  et~al.}{2021}]{DiValentino21}
{Di Valentino} E.,  et~al., 2021, \mn@doi [Classical and Quantum Gravity]
  {10.1088/1361-6382/ac086d}, \href
  {https://ui.adsabs.harvard.edu/abs/2021CQGra..38o3001D} {38, 153001}

\bibitem[\protect\citeauthoryear{{Foreman-Mackey} et~al.,}{{Foreman-Mackey}
  et~al.}{2013}]{Foreman-Mackey13}
{Foreman-Mackey} D.,  et~al., 2013, {emcee: The MCMC Hammer} (\mn@eprint {ascl}
  {1303.002})

\bibitem[\protect\citeauthoryear{{Freedman}}{{Freedman}}{2017}]{Freedman17}
{Freedman} W.~L.,  2017, \mn@doi [Nature Astronomy] {10.1038/s41550-017-0169},
  \href {https://ui.adsabs.harvard.edu/abs/2017NatAs...1E.169F} {1, 0169}

\bibitem[\protect\citeauthoryear{{Hagstotz}, {Reischke}  \& {Lilow}}{{Hagstotz}
  et~al.}{2022}]{Hagstotz21}
{Hagstotz} S.,  {Reischke} R.,   {Lilow} R.,  2022, \mn@doi [\mnras]
  {10.1093/mnras/stac077}, \href
  {https://ui.adsabs.harvard.edu/abs/2022MNRAS.511..662H} {511, 662}

\bibitem[\protect\citeauthoryear{{Heintz} et~al.,}{{Heintz}
  et~al.}{2020}]{Heintz20}
{Heintz} K.~E.,  et~al., 2020, \mn@doi [\apj] {10.3847/1538-4357/abb6fb}, \href
  {https://ui.adsabs.harvard.edu/abs/2020ApJ...903..152H} {903, 152}

\bibitem[\protect\citeauthoryear{{Jaroszynski}}{{Jaroszynski}}{2019}]{Jaroszynski19}
{Jaroszynski} M.,  2019, \mn@doi [\mnras] {10.1093/mnras/sty3529}, \href
  {https://ui.adsabs.harvard.edu/abs/2019MNRAS.484.1637J} {484, 1637}

\bibitem[\protect\citeauthoryear{{Jaroszy{\'n}ski}}{{Jaroszy{\'n}ski}}{2020}]{Jaroszynski20}
{Jaroszy{\'n}ski} M.,  2020, \mn@doi [\actaa] {10.32023/0001-5237/70.2.1},
  \href {https://ui.adsabs.harvard.edu/abs/2020AcA....70...87J} {70, 87}

\bibitem[\protect\citeauthoryear{{Katz}}{{Katz}}{2018}]{Katz18}
{Katz} J.~I.,  2018, \mn@doi [Progress in Particle and Nuclear Physics]
  {10.1016/j.ppnp.2018.07.001}, \href
  {https://ui.adsabs.harvard.edu/abs/2018PrPNP.103....1K} {103, 1}

\bibitem[\protect\citeauthoryear{{Kirsten} et~al.,}{{Kirsten}
  et~al.}{2022}]{Kirsten21}
{Kirsten} F.,  et~al., 2022, \mn@doi [\nat] {10.1038/s41586-021-04354-w}, \href
  {https://ui.adsabs.harvard.edu/abs/2022Natur.602..585K} {602, 585}

\bibitem[\protect\citeauthoryear{{Law} et~al.,}{{Law} et~al.}{2020}]{Law20}
{Law} C.~J.,  et~al., 2020, \mn@doi [\apj] {10.3847/1538-4357/aba4ac}, \href
  {https://ui.adsabs.harvard.edu/abs/2020ApJ...899..161L} {899, 161}

\bibitem[\protect\citeauthoryear{{Li}, {Gao}, {Ding}, {Wang}  \& {Zhang}}{{Li}
  et~al.}{2018}]{Li18}
{Li} Z.-X.,  {Gao} H.,  {Ding} X.-H.,  {Wang} G.-J.,   {Zhang} B.,  2018,
  \mn@doi [Nature Communications] {10.1038/s41467-018-06303-0}, \href
  {https://ui.adsabs.harvard.edu/abs/2018NatCo...9.3833L} {9, 3833}

\bibitem[\protect\citeauthoryear{{Li}, {Gao}, {Wei}, {Yang}, {Zhang}  \&
  {Zhu}}{{Li} et~al.}{2020}]{Li2020}
{Li} Z.,  {Gao} H.,  {Wei} J.~J.,  {Yang} Y.~P.,  {Zhang} B.,   {Zhu} Z.~H.,
  2020, \mn@doi [\mnras] {10.1093/mnrasl/slaa070}, \href
  {https://ui.adsabs.harvard.edu/abs/2020MNRAS.496L..28L} {496, L28}

\bibitem[\protect\citeauthoryear{{Lorimer}, {Bailes}, {McLaughlin}, {Narkevic}
  \& {Crawford}}{{Lorimer} et~al.}{2007}]{Lorimer07}
{Lorimer} D.~R.,  {Bailes} M.,  {McLaughlin} M.~A.,  {Narkevic} D.~J.,
  {Crawford} F.,  2007, \mn@doi [Science] {10.1126/science.1147532}, \href
  {https://ui.adsabs.harvard.edu/abs/2007Sci...318..777L} {318, 777}

\bibitem[\protect\citeauthoryear{{Macquart} et~al.,}{{Macquart}
  et~al.}{2020}]{Macquart20}
{Macquart} J.~P.,  et~al., 2020, \mn@doi [\nat] {10.1038/s41586-020-2300-2},
  \href {https://ui.adsabs.harvard.edu/abs/2020Natur.581..391M} {581, 391}

\bibitem[\protect\citeauthoryear{{Marcote} et~al.,}{{Marcote}
  et~al.}{2020}]{Marcote20}
{Marcote} B.,  et~al., 2020, \mn@doi [\nat] {10.1038/s41586-019-1866-z}, \href
  {https://ui.adsabs.harvard.edu/abs/2020Natur.577..190M} {577, 190}

\bibitem[\protect\citeauthoryear{{McQuinn}}{{McQuinn}}{2014}]{McQuinn14}
{McQuinn} M.,  2014, \mn@doi [\apjl] {10.1088/2041-8205/780/2/L33}, \href
  {https://ui.adsabs.harvard.edu/abs/2014ApJ...780L..33M} {780, L33}

\bibitem[\protect\citeauthoryear{{Mu{\~n}oz}, {Kovetz}, {Dai}  \&
  {Kamionkowski}}{{Mu{\~n}oz} et~al.}{2016}]{Munoz16}
{Mu{\~n}oz} J.~B.,  {Kovetz} E.~D.,  {Dai} L.,   {Kamionkowski} M.,  2016,
  \mn@doi [\prl] {10.1103/PhysRevLett.117.091301}, \href
  {https://ui.adsabs.harvard.edu/abs/2016PhRvL.117i1301M} {117, 091301}

\bibitem[\protect\citeauthoryear{{Petroff}, {Hessels}  \& {Lorimer}}{{Petroff}
  et~al.}{2019}]{Petroff19}
{Petroff} E.,  {Hessels} J.~W.~T.,   {Lorimer} D.~R.,  2019, \mn@doi [\aapr]
  {10.1007/s00159-019-0116-6}, \href
  {https://ui.adsabs.harvard.edu/abs/2019A&ARv..27....4P} {27, 4}

\bibitem[\protect\citeauthoryear{{Planck Collaboration} et~al.,}{{Planck
  Collaboration} et~al.}{2020}]{Planck20}
{Planck Collaboration} et~al., 2020, \mn@doi [\aap]
  {10.1051/0004-6361/201833910}, \href
  {https://ui.adsabs.harvard.edu/abs/2020A&A...641A...6P} {641, A6}

\bibitem[\protect\citeauthoryear{{Platts}, {Weltman}, {Walters}, {Tendulkar},
  {Gordin}  \& {Kandhai}}{{Platts} et~al.}{2019}]{Platts19}
{Platts} E.,  {Weltman} A.,  {Walters} A.,  {Tendulkar} S.~P.,  {Gordin}
  J.~E.~B.,   {Kandhai} S.,  2019, \mn@doi [\physrep]
  {10.1016/j.physrep.2019.06.003}, \href
  {https://ui.adsabs.harvard.edu/abs/2019PhR...821....1P} {821, 1}

\bibitem[\protect\citeauthoryear{{Prochaska} \& {Zheng}}{{Prochaska} \&
  {Zheng}}{2019}]{Prochaska2019}
{Prochaska} J.~X.,  {Zheng} Y.,  2019, \mn@doi [\mnras] {10.1093/mnras/stz261},
  \href {https://ui.adsabs.harvard.edu/abs/2019MNRAS.485..648P} {485, 648}

\bibitem[\protect\citeauthoryear{{Prochaska} et~al.,}{{Prochaska}
  et~al.}{2019}]{Prochaska19}
{Prochaska} J.~X.,  et~al., 2019, \mn@doi [Science] {10.1126/science.aay0073},
  \href {https://ui.adsabs.harvard.edu/abs/2019Sci...366..231P} {366, 231}

\bibitem[\protect\citeauthoryear{{Qiu}, {Zhao}, {Wang}, {Zhang}  \&
  {Zhang}}{{Qiu} et~al.}{2022}]{Qiu2021}
{Qiu} X.-W.,  {Zhao} Z.-W.,  {Wang} L.-F.,  {Zhang} J.-F.,   {Zhang} X.,  2022,
  \mn@doi [\jcap] {10.1088/1475-7516/2022/02/006}, \href
  {https://ui.adsabs.harvard.edu/abs/2022JCAP...02..006Q} {2022, 006}

\bibitem[\protect\citeauthoryear{{Ravi} et~al.,}{{Ravi} et~al.}{2019}]{Ravi19}
{Ravi} V.,  et~al., 2019, \mn@doi [\nat] {10.1038/s41586-019-1389-7}, \href
  {https://ui.adsabs.harvard.edu/abs/2019Natur.572..352R} {572, 352}

\bibitem[\protect\citeauthoryear{{Ravi} et~al.,}{{Ravi} et~al.}{2021}]{Ravi21}
{Ravi} V.,  et~al., 2021, arXiv e-prints, \href
  {https://ui.adsabs.harvard.edu/abs/2021arXiv210609710R} {p. arXiv:2106.09710}

\bibitem[\protect\citeauthoryear{{Riess}}{{Riess}}{2020}]{Riess20}
{Riess} A.~G.,  2020, \mn@doi [Nature Reviews Physics]
  {10.1038/s42254-019-0137-0}, \href
  {https://ui.adsabs.harvard.edu/abs/2020NatRP...2...10R} {2, 10}

\bibitem[\protect\citeauthoryear{{Riess}, {Casertano}, {Yuan}, {Bowers},
  {Macri}, {Zinn}  \& {Scolnic}}{{Riess} et~al.}{2021}]{Riess21}
{Riess} A.~G.,  {Casertano} S.,  {Yuan} W.,  {Bowers} J.~B.,  {Macri} L.,
  {Zinn} J.~C.,   {Scolnic} D.,  2021, \mn@doi [\apjl]
  {10.3847/2041-8213/abdbaf}, \href
  {https://ui.adsabs.harvard.edu/abs/2021ApJ...908L...6R} {908, L6}

\bibitem[\protect\citeauthoryear{{Shull}, {Smith}  \& {Danforth}}{{Shull}
  et~al.}{2012}]{Shull12}
{Shull} J.~M.,  {Smith} B.~D.,   {Danforth} C.~W.,  2012, \mn@doi [\apj]
  {10.1088/0004-637X/759/1/23}, \href
  {https://ui.adsabs.harvard.edu/abs/2012ApJ...759...23S} {759, 23}

\bibitem[\protect\citeauthoryear{{Springel} et~al.,}{{Springel}
  et~al.}{2018}]{Springel18}
{Springel} V.,  et~al., 2018, \mn@doi [\mnras] {10.1093/mnras/stx3304}, \href
  {https://ui.adsabs.harvard.edu/abs/2018MNRAS.475..676S} {475, 676}

\bibitem[\protect\citeauthoryear{{Walters}, {Weltman}, {Gaensler}, {Ma}  \&
  {Witzemann}}{{Walters} et~al.}{2018}]{Walters18}
{Walters} A.,  {Weltman} A.,  {Gaensler} B.~M.,  {Ma} Y.-Z.,   {Witzemann} A.,
  2018, \mn@doi [\apj] {10.3847/1538-4357/aaaf6b}, \href
  {https://ui.adsabs.harvard.edu/abs/2018ApJ...856...65W} {856, 65}

\bibitem[\protect\citeauthoryear{{Wang} \& {Wang}}{{Wang} \&
  {Wang}}{2018}]{Wang2018}
{Wang} Y.~K.,  {Wang} F.~Y.,  2018, \mn@doi [\aap]
  {10.1051/0004-6361/201731160}, \href
  {https://ui.adsabs.harvard.edu/abs/2018A&A...614A..50W} {614, A50}

\bibitem[\protect\citeauthoryear{{Wong} et~al.,}{{Wong}
  et~al.}{2020}]{Wong2020}
{Wong} K.~C.,  et~al., 2020, \mn@doi [\mnras] {10.1093/mnras/stz3094}, \href
  {https://ui.adsabs.harvard.edu/abs/2020MNRAS.498.1420W} {498, 1420}

\bibitem[\protect\citeauthoryear{{Wu}, {Yu}  \& {Wang}}{{Wu}
  et~al.}{2020}]{Wu2020}
{Wu} Q.,  {Yu} H.,   {Wang} F.~Y.,  2020, \mn@doi [\apj]
  {10.3847/1538-4357/ab88d2}, \href
  {https://ui.adsabs.harvard.edu/abs/2020ApJ...895...33W} {895, 33}

\bibitem[\protect\citeauthoryear{{Xiao}, {Wang}  \& {Dai}}{{Xiao}
  et~al.}{2021}]{Xiao21}
{Xiao} D.,  {Wang} F.,   {Dai} Z.,  2021, \mn@doi [Science China Physics,
  Mechanics, and Astronomy] {10.1007/s11433-020-1661-7}, \href
  {https://ui.adsabs.harvard.edu/abs/2021SCPMA..6449501X} {64, 249501}

\bibitem[\protect\citeauthoryear{{Yao}, {Manchester}  \& {Wang}}{{Yao}
  et~al.}{2017}]{Yao17}
{Yao} J.~M.,  {Manchester} R.~N.,   {Wang} N.,  2017, \mn@doi [\apj]
  {10.3847/1538-4357/835/1/29}, \href
  {https://ui.adsabs.harvard.edu/abs/2017ApJ...835...29Y} {835, 29}

\bibitem[\protect\citeauthoryear{{Yu} \& {Wang}}{{Yu} \& {Wang}}{2017}]{Yu17}
{Yu} H.,  {Wang} F.~Y.,  2017, \mn@doi [\aap] {10.1051/0004-6361/201731607},
  \href {https://ui.adsabs.harvard.edu/abs/2017A&A...606A...3Y} {606, A3}

\bibitem[\protect\citeauthoryear{{Yu}, {Ratra}  \& {Wang}}{{Yu}
  et~al.}{2018}]{Yu18}
{Yu} H.,  {Ratra} B.,   {Wang} F.-Y.,  2018, \mn@doi [\apj]
  {10.3847/1538-4357/aab0a2}, \href
  {https://ui.adsabs.harvard.edu/abs/2018ApJ...856....3Y} {856, 3}

\bibitem[\protect\citeauthoryear{{Zhang}, {Yu}, {He}  \& {Wang}}{{Zhang}
  et~al.}{2020}]{Zhang20}
{Zhang} G.~Q.,  {Yu} H.,  {He} J.~H.,   {Wang} F.~Y.,  2020, \mn@doi [\apj]
  {10.3847/1538-4357/abaa4a}, \href
  {https://ui.adsabs.harvard.edu/abs/2020ApJ...900..170Z} {900, 170}

\bibitem[\protect\citeauthoryear{{Zhang}, {Yan}, {Li}, {Zhang}  \&
  {Wang}}{{Zhang} et~al.}{2021}]{Zhangzj21}
{Zhang} Z.~J.,  {Yan} K.,  {Li} C.~M.,  {Zhang} G.~Q.,   {Wang} F.~Y.,  2021,
  \mn@doi [\apj] {10.3847/1538-4357/abceb9}, \href
  {https://ui.adsabs.harvard.edu/abs/2021ApJ...906...49Z} {906, 49}

\bibitem[\protect\citeauthoryear{{Zhao}, {Li}, {Qi}, {Gao}, {Zhang}  \&
  {Zhang}}{{Zhao} et~al.}{2020}]{Zhao2020}
{Zhao} Z.-W.,  {Li} Z.-X.,  {Qi} J.-Z.,  {Gao} H.,  {Zhang} J.-F.,   {Zhang}
  X.,  2020, \mn@doi [\apj] {10.3847/1538-4357/abb8ce}, \href
  {https://ui.adsabs.harvard.edu/abs/2020ApJ...903...83Z} {903, 83}

\bibitem[\protect\citeauthoryear{{Zheng}, {Ofek}, {Kulkarni}, {Neill}  \&
  {Juric}}{{Zheng} et~al.}{2014}]{Zheng14}
{Zheng} Z.,  {Ofek} E.~O.,  {Kulkarni} S.~R.,  {Neill} J.~D.,   {Juric} M.,
  2014, \mn@doi [\apj] {10.1088/0004-637X/797/1/71}, \href
  {https://ui.adsabs.harvard.edu/abs/2014ApJ...797...71Z} {797, 71}

\bibitem[\protect\citeauthoryear{{Zhou}, {Li}, {Wang}, {Fan}  \& {Wei}}{{Zhou}
  et~al.}{2014}]{Zhou14}
{Zhou} B.,  {Li} X.,  {Wang} T.,  {Fan} Y.-Z.,   {Wei} D.-M.,  2014, \mn@doi
  [\prd] {10.1103/PhysRevD.89.107303}, \href
  {https://ui.adsabs.harvard.edu/abs/2014PhRvD..89j7303Z} {89, 107303}

\makeatother
\end{thebibliography}


\bsp	
\label{lastpage}
\end{document}